\documentclass[submission,copyright,creativecommons]{eptcs}
\providecommand{\event}{ICLP 2021 Technical Communications} 

\usepackage{mathptmx}
\usepackage{amsmath}
\usepackage{algorithm}
\usepackage{graphicx}
\usepackage{algpseudocode}

\usepackage{listings}

\newtheorem{definition}{Definition}

\usepackage{comment}
\usepackage{cprotect}

\usepackage{fancyvrb}
\usepackage{times}
\usepackage{soul}
\usepackage{url}
\usepackage{booktabs}
\usepackage{subcaption}

\title{DiscASP: A Graph-based ASP System for Finding Relevant Consistent Concepts with Applications to Conversational Socialbots}
\author{Fang Li, Huaduo Wang, Kinjal Basu, Elmer Salazar, Gopal Gupta
\institute{University of Texas at Dallas\\ Richardson, USA}
\email{$\{$fang.li, huaduo.wang, kinjal.basu, elmer.salazar, gupta$\}$@utdallas.edu}
}
\def\titlerunning{DiscASP}
\def\authorrunning{F. Li, H. Wang, K. Basu, E. Salazar \& G. Gupta}
\begin{document}
\maketitle

\begin{abstract}
We consider the problem of finding \textit{relevant consistent concepts} in a conversational AI system, particularly, for realizing a conversational socialbot. Commonsense knowledge about various topics can be represented as an answer set program. However, to advance the conversation, we need to solve the problem of finding \textit{relevant consistent concepts}, i.e., find consistent knowledge in the ``neighborhood" of the current topic being discussed that can be used to advance the conversation. Traditional ASP solvers will generate the whole answer set which is stripped of all the associations between the various atoms (concepts) and thus cannot be used to find relevant consistent concepts. Similarly, goal-directed implementations of ASP will only find concepts \textit{directly} relevant to a query. We present the DiscASP system that will find the partial consistent model that is relevant to a given topic in a manner similar to how a human will find it. DiscASP is based on a novel graph-based algorithm for finding stable models of an answer set program. We present the DiscASP algorithm, its implementation, and its application to developing a conversational socialbot.
\end{abstract}

\section{Introduction}

Conversational AI has been an active area of research, starting from a rule-based system, such as ELIZA \cite{eliza} and PARRY \cite{parry}, to the recent open domain, data driven conversational agents like Amazon's Alexa \cite{caspr}. Early rule-based bots were based on just syntax analysis and thus were limited, while the main challenge of modern machine learning (ML) based chatbots is the lack of ``understanding'' of the dialogs in the conversation. Current machine learning technology-based chat-bots \cite{chatbot-ref} learn patterns in data from large corpora and compute a response without having any semantic understanding of the conversation utterances. Such an approach limits these chatbots, e.g.,  they may give an irrelevant response or they cannot provide an explanation for their response. We believe that a chatbot that can converse like a human \textit{has to} employ semantic understanding and reasoning, aside from other AI technologies such as machine learning and natural language processing.

The goal of our research is to develop a \textit{conversational socialbot} that can hold a conversation with a human stranger on topics of general interest such as movies, books, music, pets, family, etc. Our work is inspired by the Amazon Alexa Socialbot Challenge competition \cite{alexa,casprutd} in which the authors' participated. 
A realistic socialbot should be able to understand and reason like a human. In human to human conversations, we do not always tell every detail, we expect the listener to fill gaps through their commonsense knowledge. Thus, modeling commonsense knowledge and commonsense reasoning is an important consideration in developing a socialbot.

As is well known, the human thought process is flexible and non-monotonic in nature, which means \textit{what we believe now may become false in the future with new knowledge}. Given the need for nonmonotonic reasoning, we use Answer Set Programming (ASP) \cite{gelfond2014knowledge} as the underlying formalism for commonsense knowledge representation and reasoning. With the use of (i) default rules, (ii) exceptions to defaults, and (iii) preferences over multiple defaults, one can model bulk of human-style reasoning \cite{gelfond2014knowledge}, at least those parts that are needed for realizing a socialbot. 

One of the problems that need to be addressed in developing a conversational socialbot is inferring the knowledge ``relevant" to the topic at hand. That is, concepts that are consistent with the knowledge we possess about the topic. We call these concepts \textit{relevant consistent concepts} for that topic.
During a conversation about movies, for example, if the other person says that their favorite movie is Titanic, then we immediately will try to recall all the knowledge we possess about that topic. Thus, we may remember that Titanic's main actors were Leonardo Di Caprio and Kate Winslett, that Titanic got many Academy awards, and that the director of the movie, James Cameron, became very famous. We may then advance the conversation by saying, ``Yes, Titanic was a wonderful movie, and it got many Academy Awards."  
Social conversations drift from a topic to another related topic. For example, after talking a little bit more about Titanic, the conversation may shift to another movie directed by Titanic's director James Cameron such as Avatar. Or, it may shift to another movie of one of the actors in Titanic, e.g., Wolf of Wall Street, in which Leonardo Di Caprio was also the lead actor. This knowledge that we recall about Titanic is the  \textit{relevant consistent concept} (RCC) for Titanic. Finding the RCC, given a topic, is an important problem in developing conversational socialbots, as it helps in moving the conversation along. In this paper, we focus on the problem of efficiently finding the RCC({$\tau$}) for a topic atom {$\tau$}, given a commonsense knowledgebase coded in ASP. In this paper, we restrict ourselves to conversations about movies, though the approach applies to any domain. 

We present an efficient graph-based algorithm for computing the RCC({$\tau$}) for a topic {$\tau$} given relevant commonsense knowledge coded in ASP. The program can only have headless constraints, while constraints realized through \textit{odd loop over negation} are not permitted. This is done for simplicity: such restricted programs are sufficient for modeling conversational socialbots.
Our algorithm computes the set of atoms that represent the concepts that are related to the topic at hand in a given answer set. This set of atoms is computed incrementally, i.e., it is not based on computing the entire answer set. 
Given a query, we want to not only find out a justification for it, but also the related associated knowledge. E.g., given a query {\small {\tt ?-  flies(tweety)}}, if we infer that Tweety flies because Tweety is a bird due to the rule {\small {\tt flies(X) :- bird(X), not ab\_bird(X)}}., then we also want to know the associated concept that Tweety has wings from the rule {\small {\tt haswings(X) :- bird(X)}}. This is very similar to how humans will bring in relevant concepts in their current working memory through a combination of backward and forward chaining.

Our algorithm (called DiscASP) guarantees that the computed partial answer set is part of a complete answer set dictated by the Gelfond-Lifschitz transform \cite{gelfond2014knowledge}. Traditional tools for ASP such as CLINGO \cite{clingo} are not suitable for this task, as they compute the entire answer set. The entire answer set can be quite large as the amount of commonsense knowledge can be enormous. Even if we restrict ourselves to conversations about movies, the movie database that a socialbot may have to work with will have information about at least 50 movies (assuming that a person can only keep information about that many movies in their head). In reality, this number can be much larger (e.g., the iMDB movie database has more than 7.5 Million entries). In addition, systems such as CLINGO lose all the information connecting two atoms during the grounding process. Thus, explaining the connections as we leap from topic to topic in a conversation is difficult.

Our contribution is twofold: (i) we present an efficient graph-based algorithm for computing a subset of atoms of an answer set that are consistent with, and relevant to, a topic atom present in the answer set. To the best of our knowledge, the algorithm is novel. The algorithm also provides the logical relationship between the topic atom and any other atom in the computed partial answer set based on the commonsense knowledge rules provided; (ii) we present a practical way of solving the important problem of computing the set of relevant consistent concepts for a  topic atom while developing conversational socialbots. 

%

\section{Background}

Answer set programming (ASP) \cite{MT5} is a popular nonmonotonic-logic based paradigm for knowledge representation and reasoning and for solving combinatorial problems. 
Most ASP solvers employ SAT solver-like technology to find these answer sets. As a result, justification for why a literal is in the answer set is hard to produce. Compared to SAT solver based implementations,  graph-based implementations of ASP have not been well studied. 
%
Graph-based approaches for ASP \cite{anger2001nomore,konczak2005graphs,linke2005suitable} are well designed, but their graph representations are complex as they all rely on extra information to map the ASP elements to nodes and edges of a graph. In contrast, our approach uses a much simpler graph representation, where nodes represent literals and an edge represents the relationship between the nodes it connects. Since this representation faithfully reflects the causal relationships, it is capable of producing causal justification for goals entailed by the program. In this section, we introduce the background concepts of our novel graph representation.

\medskip\noindent\textbf{Call Dependency Graph:}\label{sec:dg}
A call dependency graph or dependency graph \cite{linke2005suitable} uses nodes and directed edges to represent call dependency relationships in an ASP rule. 

\begin{definition}
The dependency graph of a program is defined on its literals s.t. there is a positive (resp. negative) edge from $q$ to $p$ if $q$ appears positively (resp. negatively) in the body of a rule with head $p$.
\label{def1}
\end{definition}

\begin{figure}[tb]
\centering
\begin{minipage}{.5\textwidth}
  \centering
  \includegraphics[scale=0.34]{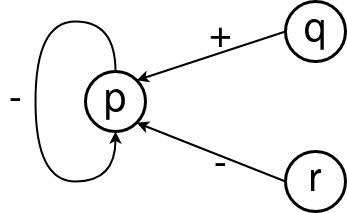}
      \caption{Dep. Graph for Programs 1 \& 2}
    \label{fig:fig1}

\end{minipage}%
\begin{minipage}{.5\textwidth}
  \centering
    \begin{subfigure}[b]{0.5\linewidth}
        \centering
        \includegraphics[scale=0.33]{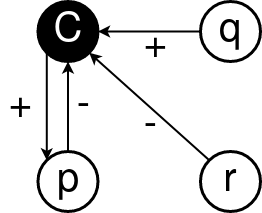}
        \caption{CNR for Program 1}
    \end{subfigure}\hfill
    \begin{subfigure}[b]{0.5\linewidth}
        \centering
        \includegraphics[scale=0.33]{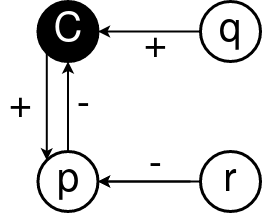}
        \caption{CNR for Program 2}
    \end{subfigure}
    \caption{CNRs for Program 1 \& 2}
    \label{fig:fig2}
\end{minipage}
\end{figure}

Conventional dependency graphs are not able to represent ASP programs uniquely. This is due to the inability of dependency graphs to distinguish between non-determinism (multiple rules defining a proposition) and conjunctions (multiple conjunctive sub-goals in the body of a rule) in logic programs. For example, the following two programs have identical dependency graphs (Figure \ref{fig:fig1}).

\noindent\begin{minipage}{.45\textwidth}
\begin{lstlisting}[language=prolog,basicstyle=\footnotesize]
    %% program 1  
    p :- q, not r, not p.
\end{lstlisting}
\end{minipage}
\begin{minipage}{.45\textwidth}
\begin{lstlisting}[language=prolog,basicstyle=\footnotesize]
    %% program 2
    p :- q, not p. p :- not r.
\end{lstlisting}
\end{minipage}

To make conjunctive relationships representable by dependency graphs, we first transform them slightly to come up with a novel representation method. This new representation method, called conjunction node representation (CNR) graph, uses an artificial conjunction node to represent a conjunction of sub-goals in the body of a rule (Figure \ref{fig:fig2}). The conjunction node (colored in black) refers to the conjunctive relation between the incoming edges from nodes representing sub-goals in the body of a rule. 
Note that a CNR graph is not a conventional dependency graph.

\begin{figure}[tb]
\centering
\begin{minipage}{.5\textwidth}
  \centering
  \includegraphics[width=.6\linewidth]{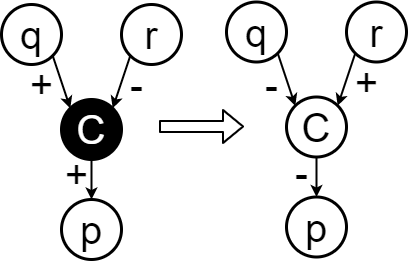}
  \caption{CNR-DG Transformation}
  \label{fig:fig3}
\end{minipage}%
\begin{minipage}{.5\textwidth}
  \centering
    \begin{subfigure}[b]{0.5\linewidth}
        \centering
        \includegraphics[scale=0.3]{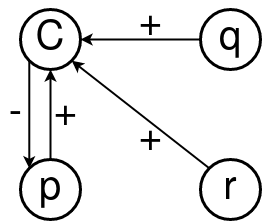}
        \caption{Program 4}
        \label{fig:fig4a}
    \end{subfigure}\hfill
    \begin{subfigure}[b]{0.5\linewidth}
        \centering
        \includegraphics[scale=0.3]{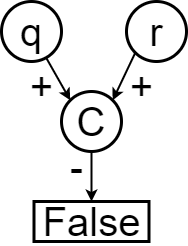}
        \caption{Program 5}
        \label{fig:fig4b}
    \end{subfigure}
    \caption{Constraint DG}
    \label{fig:fig4}
\end{minipage}
\end{figure}

\medskip\noindent\textbf{Converting CNR Graph to Dependency Graph:} \label{sec:cnrtodg}
Since CNR graph does not follow the dependency graph convention, we need to convert it to a proper dependency graph in order to perform dependency graph-based reasoning. We use a simple technique to convert a CNR graph to an equivalent conventional dependency graph. We negate all in-edges and out-edges of the conjunction node. This process essentially converts a conjunction into a disjunction. As an example, Figure \ref{fig:fig3} shows the CNR graph to dependency graph transformation for Program 3. We call such a dependency graph a CNR Dependency Graph. 
The transformation is a simple application of De Morgan's law. The rule in program {\small {\tt p :- q, not r.}} can be represented as: {\small {\tt p :- conj. conj :- q, not r.}} The transformation produces the equivalent rules: {\small {\tt p :- not conj. conj :- not q. conj :- r.}} Since conjunction nodes are just helper nodes that allow us to perform dependency graph reasoning, we don't report them in the final answer set. The transformation process is quite straightforward, so we do not give any more details.

ASP also allows for special types of rules called constraints. There are two ways to encode constraints: (i) headed constraint where the negated head is called directly or indirectly in the body (e.g., Program 3), and (ii) headless constraints (e.g., Program 4). 

\noindent\begin{minipage}{.45\textwidth}
\begin{lstlisting}[language=prolog,basicstyle=\footnotesize]
    %% program 3
    p :- not q, not r, not p.
\end{lstlisting}
\end{minipage}
\begin{minipage}{.45\textwidth}
\begin{lstlisting}[language=prolog,basicstyle=\footnotesize]
    %% program 4
    :- not q, not r.
\end{lstlisting}
\end{minipage}

Our algorithm models these constraint types separately. For the former one, we just need to apply the CNR-DG transformation directly. Note that the head node connects to the conjunction node both with an in-coming edge and an out-going edge (Figure \ref{fig:fig4a}). For the headless constraint, we create a head node with the truth value as \textit{False}. For simplicity, we currently only permit headless constraints, as these are sufficient to model conversational socialbots. 

The reason why we don't treat a headless constraint the same way as a headed constraint is because in the latter case, if head node ({\tt p} in Program 3) is provable through another rule, then the headed constraint is inapplicable. Therefore, we cannot simply assign a false value to its head. 

\section{A Graph Algorithm for Computing Partial Answer Sets} \label{sec:algorithm}

We have developed DiscASP, a graph-based algorithm, for finding partial answer sets. The philosophy of DiscASP is to translate an ASP program into a CNR dependency graph, which is always constrained by some constraint rules, then try to satisfy the constraints by assigning presumed truth values to the related nodes, until all constraints have been satisfied. At the same time, DiscASP will propagate truth values of the nodes whose truth value has already been determined.
Our graph-based approach performs reasoning in an incremental manner. It starts from the constraints in the answer set program and traces along with causal nodes until it finds support through facts (well-founded case) or it detects a cycle through negation (cyclic case). Our algorithm can be thought of as a more general form of the Galliwasp algorithm for query-driven execution of answer set programs \cite{galliwasp2}. The DiscASP approach is constraint-driven and thus significantly reduces the search space by avoiding the exploration of worlds that are inconsistent with the constraints. Furthermore, the incremental reasoning from constraints allows DiscASP to perform query-driven execution. 

\medskip
\noindent\textbf{Input:} 
At present, the DiscASP algorithm takes only pure grounded propositional ASP programs as input. A valid rule should be in the form of $head$ :- $body.$, :- $body.$, or $head.$ For example, if we want to represent 3 balls, the form $ball(1..3)$ is invalid. Instead, we have to declare them separately as $ball(1).$ $ball(2).$ $ball(3).$ 
The input ASP program will be converted and saved into a directed-graph data structure representing the CNR dependency graph. The conversion process is based on the concepts that were introduced in Section \ref{sec:dg}. When given a query {\small {\tt Q}}, the additional constraint {\small {\tt :- not Q}} will be appended to the original ASP program. 

\begin{figure}[tb]
    \centering
    \includegraphics[scale=0.30]{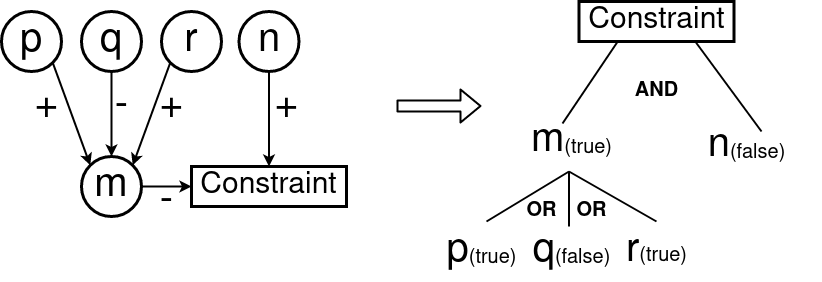}
    \caption{Satisfiability Example}
    \label{fig:incre-graph}
\end{figure}

\medskip\noindent\textbf{The DiscASP Algorithm:} \label{sec:igASPalgorithm}
The DiscASP algorithm is a recursive algorithm. Since a CNR dependency graph represents the causal relationships among nodes, a topological order would indicate the truth values flow along edges from one node to another, starting from the leaves. By their nature, the constraint nodes (labeled \textit{False}, discussed in Section \ref{sec:dg}) will be at the end of such flows in the CNR dependency graph. Therefore, we can incrementally establish the satisfiability relationships across all the nodes starting from the constraint nodes. This incremental establishment of satisfiability starting from the constraint nodes amounts to developing a proof tree. 
\begin{lstlisting}[language=prolog,basicstyle=\footnotesize]
    %% program 5
    m :- p.      m :- not q.      m :- r.      :- not m.     :- n.
\end{lstlisting}
An example (Program 5) is shown in Figure \ref{fig:incre-graph} where the graph is to the left and the proof tree to the right. To falsify the constraint node, i.e., to ensure it is \textit{False}, node $m$ must be \textit{True} and $n$ must be \textit{False}. For node $m$ to be \textit{True}, at least one of the three must hold: $p$ is \textit{True}, $q$ is \textit{False}, or $r$ is \textit{True}. When every node's presumed truth value has been found to be consistent with all the dependencies, the algorithm will return the answers. Algorithm \ref{alg} shows an abstraction of the core algorithm. Here we explain some key concepts as follows:

\begin{algorithm}[tb]
\small
	\caption{DiscASP core algorithm (abstraction)}
	\begin{algorithmic}[1]
	\Procedure{reasoningRec}{state, node, value}

	    \If {(value == false)} \Comment{case 1: presume the current node to be false}
    		\If{the current node's presumed value can be proved}
    		    \State add the current node to facts according to its proved value
    		    \State propagate truth values
    		    \State return facts
    	    \Else
    	        \For {each predecessor}
    	            \State assign truth value which doesn't make the current node true
    	            \State recursively find the sub-answer set accordingly
    	        \EndFor
    		    \State run \textbf{conjunctive} merge
    		    \State \textbf{return} the results
    	    \EndIf
		
	    \Else \Comment{case 2: presume the current node to be true}
    		\If{the current node's presumed value can be proved}
    		    \State add the current node to facts according to its proved value
    		    \State propagate truth values
    		    \State return facts
    	    \Else
    	        \For {each predecessor}
    	            \State assign truth value which doesn't make the current node false
    	            \State recursively find the sub-answer set accordingly
    	        \EndFor
    		    \State run \textbf{disjunctive} merge
    		    \State \textbf{return} the results
    		\EndIf
	    \EndIf
	\EndProcedure
	\end{algorithmic}
	\label{alg}
\end{algorithm}

\medskip\noindent\textbf{Effective Edge:}
An effective edge in a CNR dependency graph refers to any edge that propagates \textit{True} value to the node it is incident on. There are two types of \textit{effective} edges: (i) positive edge emanating from a \textit{True} node; (ii) negative edge emanating from a \textit{False} node. An effective edge only points to a \textit{True} node.

\medskip\noindent\textbf{Satisfying Conjunction vs. Disjunction:}
There are two kinds of dependencies that may arise for a node in the CNR dependency graph: conjunctive and disjunctive. A conjunctive dependency refers to the situation where a node is presumed to be \textit{False}. In this case, none of the edges incident into the node should be effective edges. A disjunctive dependency indicates that when a node is presumed to be \textit{True}, at least one of the in-coming edges should be effective. Since DiscASP works in a reverse manner (from constraints to facts), we may get multiple partial models before we can validate the \textit{True}/\textit{False} label of the current node. For both conjunction and disjunction, these partial models need to be merged for the sake of integrity as well as efficiency. The merging process is discussed later.

\medskip\noindent\textbf{Proof Branch:}
In the DiscASP algorithm, we start from the constraints that have to be shown to be false and incrementally construct a proof tree obeying the constraints imposed by the CNR dependency graph. In this incremental reasoning  process, we will pursue various paths in the CNR dependency graph. Our proof will have multiple branches, corresponding to various paths in the CNR dependency graph. Traversal of a branch stops when we reach a fact node whose value has already been given by the ASP program (i.e., known to be true due to being a fact, or known to be false because the atom does not have a rule with that atom as head), or sense that the branch contains a \textbf{loop} (discussed next).

\medskip\noindent\textbf{Loop Detection and Handling:}
In an ASP program, loops among literals may exist. There are three kinds of loops that can be found in the program: even loops, odd loops, and positive loops. Even loops and odd loops refer to loops that have an even or odd number of negative edges in the corresponding dependency graph. Positive loops are loops with no negative edge. It is well known that even loops generate multiple worlds, while odd loops kill worlds. For example, in program {\small {\tt p :- not q. q :- not p.}}, $p$ and $q$ form an even loop, which generates two mutually exclusive worlds: \{p/True, q/False\} and \{q/True, p/False\}. For program {\small {\tt p :- not q. q :- not r. r :- not p.}}, nodes $p$, $q$ and $r$ form an odd loop, which makes the program unsatisfiable. Odd loops are currently disallowed in DiscASP, as constraints are expressed through the headless constraint construct.

For positive loops, we need to ensure that any models computed are consistent with ASP semantics. Suppose we have a program {\small {\tt p :- q. q :- p.}}, under ASP semantics, it will have only one answer set:  \{p/False, q/False\}. The other model (\{p/True, q/True\}) has to be rejected, as it is not well-founded per ASP semantics. Thus, positive loops have to be handled properly so that only correct answer sets are reported. 
To detect a loop, DiscASP keeps track of the presumed nodes along the branch, when the current node has been seen previously, we will check whether there exists any \textit{False} node between these two nodes. If so, it is an even loop, otherwise, it is a positive loop and only the falsifying assignment should be computed.

\medskip\noindent\textbf{Model Merging:} 
As mentioned previously, for each presumed node $n$ (i.e., a node assigned a truth value), its dependencies will be either conjunctive (if $n$ is presumed \textit{False}) 
or disjunctive.
(if $n$ is presumed \textit{True}).
For both conditions, we need to merge the partial models that have been computed so far while assigning a truth value to the dependent nodes.
For the conjunctive condition, the merging process only takes successfully merged models, each of which is the union of two non-conflicting sub-models. 
For example, consider a node $n$ that is presumed to be \textit{False}. Suppose it has two predecessors $p$ and $q$, both $p$ and $q$ connect to $n$ via negative edges. So $n$ will only be \textit{False} when both $p$ and $q$ will be \textit{True}. We need a conjunctive merge here. Suppose we have sub-models \{p$_1$:\{a/True, d/True, b/False\}, p$_2$:\{a/False, b/True\}\} that hold for $p$ to be \textit{True}, and sub-models \{q1:\{a/True, c/True, b/False\}\} for $q$ to be \textit{True}. The conjunctive merging of sub-models between $p$ and $q$ will only accept the union of $p_1$ and $q_1$, because $p_2$ conflicts with $q_1$. Therefore, there will only be one model to satisfy for $n$ being \textit{False}, that is \{a/True, c/True, d/True, b/False\}.

For disjunctive merging, we will keep the conflicted sub-models along with successfully merged ones. Let's modify the above example a little bit by presuming the value of node $n$ to be \textit{True}, and keep everything else unchanged. Now the merging condition becomes disjunctive because one of $p$ or $q$ being \textit{False} will still make $n$ \textit{True}. Since $p_1$ and $q_1$ can be merged without conflict, we replace them with their union \{a/True, c/True, d/True, b/False\}. But this time we don't discard $p_2$, because $p_2$ is also a valid model that makes $n$ \textit{True}. Therefore, after this merging, we will have two sub-models for $n$ being \textit{True}: \{\{a/True, c/True, d/True, b/False\}, \{a/False, b/True\}\}.

\medskip\noindent\textbf{Forward Propagation:}
Since nodes are assigned values is in a backward chaining manner, where we compute the truth assignment of the predecessors before that of the current node, the sub-models need to cover as much information as possible. If some nodes' values can be inferred from the proven nodes, they must also be added to the sub-model. For example, suppose we have a sub-model \{a/True, b/False\} for making node $n$ \textit{True}. Suppose there are two additional rules related to nodes $a$ and $b$: {\small {\tt (i) c :- a. (ii) d :- not b.}} In this case, we know that $c$ and $d$ must also be \textit{True}. 

DiscASP propagates truth values every time a presumed node value has been established, by using a causal map that covers all of the causal relationships for each node/value. When a presumed node/value is established, DiscASP will check whether there is any other node whose value can be inferred from current node assignments. If there are any, the inferred value is assigned to that node and propagation continues until the model does not change.

\medskip\noindent\textbf{Query Handling:}
A query w.r.t. an ASP program amounts to checking whether a literal is in one of the models of the program. For instance, ASP program {\small {\tt p :- not q. q :- not p. :- p, q.}} has two models \{\{p/True, q/False\}, \{p/False, q/True\}\}. If we query $p$, we should get the model $\{p\}$. 

For query handling, DiscASP negates the query literal and appends it to the ASP program as an additional constraint. So for the above example, the query $p/True$ will be converted to a constraint rule {\small {\tt :- not p.}} and added to the original program. So the program will now be {\small {\tt p :- not q. q :- not p. :- p, q. :- not p.}}

DiscASP begins its reasoning from a constraint node (typically, the query represented as a constraint), then searches for a partial answer set to satisfy the constraint. 


\medskip\noindent\textbf{Theorem:}\label{theorem}
Let $P$ be a program, and $A$ be an answer generated by DiscASP for $P$. Assuming $P$ is consistent, $A$ is a subset of some stable model of $P$.

\noindent\textbf{Proof Sketch:} All constraint (headless rules) bodies are connected via incoming edges to a special constraint node. If the body of any constraint rule is \textit{True}, then the constraint will be \textit{True}. We want the constraint to be \textit{False}, and so DiscASP begins constructing answers by trying to prove that the constraint is \textit{False}. We can show by induction on the depth of the proof tree that if an answer is produced it is a subset of some stable model. 
(A detailed proof can be found in \url{http://utdallas.edu/~gupta/discasp.pdf}).

\section{Application of DiscASP to Conversational Socialbots}
In some application scenarios, finding the whole answer sets may be overkill or impractical. Especially for commonsense reasoning, where the knowledgebases may be large and guaranteeing consistency may be hard as parts of the knowledgebase were constructed separately. If a subset of the knoweldgebase that contains the answer we are seeking is consistent, we may not care about other inconsistent part of the knowledgebase. 

\medskip\noindent\textbf{Relevant Consistent Concepts:}
When we hold a social conversation, such as when we meet another person at a social event, the conversation will start from a topic such as one person asking the other ``Have you seen any movies lately?" Once the second person's response is known (e.g., the person answers Titanic), a reasonable conversation strategy is to engage the other person in a conversation around this topic. Based on our commonsense knowledge, we will immediately attempt to recall all the information we know about the movie Titanic that is consistent with and relevant to our knowledge that the other persons like it. This information may include the movie's actors, the director, trivia about the movie, awards the movie earned, the movie plot, etc. This information is recalled based on our commonsense knowledge that a person interested in a movie is also interested in its actors, director, related trivia, movie plot, awards it won, etc. and so recalling the information will help us prepare a response. We will then pick one of the aspects and respond, in order to continue the conversation. For example, we may say, that ``Oh, yes, Leonardo Di Caprio, did a great job as the lead actor." The conversation may revolve around Titanic, but may change to another movie that Leonardo Di Caprio has also acted in, such as Wolf of Wall Street. Conversations will continue to evolve from the topic of Titanic movie in this manner, until the topic changes completely, say, to pet ownership. Conversations will then ensue on this topic in a manner similar to the topic of movies. 

For a socialbot to hold such casual conversations with a human, it needs to figure out the knowledge related to a topic (represented as a propositional atom, e.g., like(john, titanic)). We term the conceptual knowledge related to a topic {$\tau$} as \textit{relevant consistent concepts} (RCC) of {$\tau$}. These concepts are inferred by humans from commonsense rules that they learn over time, for example, we might learn that \textit{normally, a person who likes a movie, will also like the actors in the movie}. Thus, if we know that John likes Titanic, we infer that very likely, he likes Leonardo Di Caprio, the lead actor, too.

\medskip\noindent\textbf{Why A Partial Answer Set Solver?:}
Finding the complete answer set of a program requires considering the entire dependency graph, which may include sub-graphs that are disconnected from each other. For example, knowledgebase about books will be largely independent of the knowledgebase about movies. A traditional answer set solver will generate answer sets where a truth assignment is generated for every atom. However, for applications such as conversational socialbot, given a topic {$\tau$}, we only want to generate the set of \textit{relevant consistent concepts} of {$\tau$}. Since the knowledge base contains more knowledge than necessary related to a topic, for example, when we hold a conversation about movies, finding the whole answer set is overkill. In our work, we use DiscASP (for \textbf{Disc}ussions with \textbf{ASP}) for computing a partial answer set representing the {\tt RCC($\tau$)}, given a topic atom {$\tau$}.

In DiscASP, if a dependency sub-graph is disconnected to a query related to a topic {$\tau$}, then for obvious reasons, the atoms in this sub-graph do not have to be considered at all to compute the {\tt RCC($\tau$)} set. Given a query for a topic {$\tau$}, we only need to verify its validity and find out its related consistent concepts. As discussed (Section \ref{theorem}), DiscASP guarantees the following property: given an answer set program {\tt P}, if {\tt P} has an answer set, then any partial answer set computed by DiscASP will be a subset of a complete answer set of {\tt P}. 

DiscASP is designed as a system that finds partial stable models for a given topic with respect to a specific knowledgebase coded in ASP.
The knowledgebase will be encoded as propositional logic facts and will contain both generic data that is collected from data sources, e.g., iMDB movie database, user profiles, etc., as well as commonsense knowledge rules that capture the knowledge of the domain. These facts and rules are expressed in ASP. The commonsense rules are grounded using the GRINGO grounding system \cite{clingo}. Since DiscASP is a graph based ASP solver, we need to transform the grounded program into a CNR dependency graph as discussed earlier. While building this graph, the topic query, {\tt Q}, will also be added as the constraint ({\tt :- not Q.}). Then the dependency graph will be passed to the partial answer set solver, which computes partial stable models. The DiscASP algorithm is quite simple: it starts with the constraint introduced by the query ({\tt :- not Q.}) and finds the smallest model, {\tt M}, that ensures that the constraint holds. If there are atoms in {\tt M} that also involve other constraints, then the {DiscASP} algorithm ensures that these constraints are also satisfied. The partial model found is the relevant consistent concept (RCC) set of {\tt Q}.

\medskip\noindent\textbf{DiscASP Illustration:}
Let's use a simple propositional ASP programs to demonstrate how DiscASP works. Consider \textit{Program 6} with query {\small {\tt ?- p}}, the full answer sets are {\small {\tt \{r, q, p, s, t, j, m, k, n, o, w, a\}}}
and {\small {\tt \{r, q, p, s, t, j, m, k, n, o, w, b\}}}. There are two answer sets 
because nodes {\tt a, b} constitute an even loop, which creates two mutually exclusive worlds. 
\begin{lstlisting}[language=prolog,basicstyle=\small]
 %% program 6
 p :- q, r.  q :- s, not x. t :- s. j :- r. m :- t. k :- j. n :- p. o :- n.
 r :- not u, not v. w :- not v. a :- not b. b :- not a. s.
\end{lstlisting}

\noindent DiscASP returns one single partial answer set {\small {\tt \{r, q, p, s, t, j, m, k, n, o, w, not x, not u, not v\}}}. Because nodes {\small {\tt a, b}} have no connection to the sub-graph that contains the query node {\tt p}, 
DiscASP only returns the most general answer containing the query node. If our query is extended to {\small {\tt ?- p, a}}, then {\small {\tt a, not b}} will be added to this answer set. 
 
\medskip\noindent\textbf{Narrowing the RCC:}
The {DiscASP} algorithm provides the facility to narrow the RCC set even further. This is because even a partial answer set may still contain too many atoms. Suppose our socialbot is chatting with a user John about movies and comes to know that the user likes the movie \textit{Titanic}. DiscASP will take {\small {\tt like\_movie(john, titanic)}} as a query, compute a partial stable model from the knowledge base consisting of the iMDB movie database. The answer set may contain atoms that are at quite some distance to the query in the CNR dependency graph. DiscASP thus also takes a radius parameter (a positive number), {\tt r}, and finds all atoms in the RCC that are at a distance {\tt r} or less from the query constraint in the CNR dependency graph. The DiscASP algorithm will also output the path between the query constraint and an arbitrary atom in the RCC. The path describes the connection between the two atoms and can be used to construct the socialbot's natural language response.
 
\begin{definition}
The set of relevant consistent concepts for a given topic {$\tau$} and radius $r$, denoted {\tt RCC($\tau$,$r$)}, is a subset of the partial answer set containing {$\tau$}, where for each atom $a$ in this partial answer set, the distance between $a$ and the atom {$\tau$} in the CNR dependency-graph is less than or equal to $r$.
\end{definition}

 
\noindent For example, from the atom {\small {\tt like\_movie(john, titanic)}}, we know that John may also like to talk about Matt Damon because he and Leonardo Di Caprio, the hero of Titanic, were together in the movie \textit{The Departed}. Then John may also like to talk about Tom Hanks who was a co-actor with Matt Damon in the movie \textit{Saving Private Ryan}. Here if we set the propagation radius to 2, DiscASP will only give us {\small {\tt like\_actor(john, matt\_damon)}}, omitting Tom Hanks. Then, our socialbot will start talking about Matt Damon. If we switch from \textit{Titanic} directly to Tom Hanks, the user may feel confused. Thus, it is important to keep the radius small to keep the topics relevant.
For program 6 above, if we set the radius to 2, for the query {\tt q}, we will compute the partial answer set {\small {\tt \{r, q, p, s, t, j, not x\}}}. 

\medskip\noindent\textbf{An Example: Socialbot Conversation about Movies:} 
\label{sec:example}
%
We have chosen the real-world example of the movie domain to demonstrate our DiscASP algorithm. First, we need to define the  ASP rule-set that captures our commonsense knowledge for holding conversations about movies. 
These rules are written in a way that portrays the human thinking process. In other words, these rules mimic the reasoning process of how a human will choose what to talk next about a particular movie in a conversation. 
For example, humans will assume that \textit{normally, if someone likes an actor, he/she likes the movies by the actor and vice-versa}. This condition can be depicted using the following two default rules with exceptions, where person \textit{P} likes a movie \textit{M} or an actor \textit{A}. The rules are self-explanatory, so to save space, details are omitted.
\begin{lstlisting}[language=prolog,basicstyle=\footnotesize]
    like_movie(P, M) :- movie(M), actor(A), like_actor(P, A), movie_actor(M, A), 
            not ab_like_movie(P, M).
    like_actor(P, A) :- movie(M), actor(A), like_movie(P, M), movie_actor(M, A),
            not ab_like_actor(P, A).
\end{lstlisting}

We use the {\tt talk\_preference/3} predicate to capture the attribute A, that the socialbot can talk about next, regarding a movie M, given a user P. For example, \textit{people prefer to talk about an actor of a movie if he/she is the main actor, a famous actor or an Oscar winning actor}. Also, \textit{normally, people like to talk about a movie's awards, awards to actors in the movie, or trivia about a movie}. Note that these default rules include exceptions, for example, to capture the situation that whenever the socialbot is done talking about an attribute it adds it to its exception list so as to not talk about it again in the same conversation. The {\small {\tt ab\_talk\_preference/3}} predicate models these exceptional situations (in the code below, the user John has already talked about the movie Avatar).
\begin{lstlisting}[language=prolog,basicstyle=\footnotesize]
    talk_preference(P, M, A) :- actor(A), movie(M), like_movie(P, M), 
            main_actor(A, M), not ab_talk_preference(P, M, A).
    talk_preference(P, M, A) :- actor(A),like_movie(P, M),movie_actor(M, A), 
            famous_actor(A), not ab_talk_preference(P, M, A). 
    talk_preference(P, M, A) :- actor(A),like_movie(P, M),movie_actor(M, A),
            award_won(A,oscar), not ab_talk_preference(P,M,A).
    talk_preference(P, M, awards) :- movie(M), like_movie(P, M), 
            not ab_talk_preference(P, M, awards).
    talk_preference(P, M, actor_award) :- movie(M), movie_actor(M, A), 
            like_actor(P, A), not ab_talk_preference(P, M, actor_award).
    talk_preference(P, M, trivia) :- movie(M), like_movie(P, M), 
            not ab_talk_preference(P, M, trivia).
    ab_talk_preference(P, M, A) :- already_talked(P,M, A).
    already_talked(john, avatar, actor).  
\end{lstlisting}

\noindent Also, we will have default rules that model knowledge about people's preferences such as: \textit{Normally, young people like action and sci-fi movies}, or\textit{ children like kid's movies}, or \textit{women like romance and drama movies}. One example rule is given below.
\begin{lstlisting}[language=prolog,basicstyle=\footnotesize]
    like_movie(P, M) :- movie(M), age_category(P, young), genre(M, scifi), 
            not ab_like_movie(P, M).
\end{lstlisting}            

\noindent Human reasoning process utilizes \textit{constraints} on a regular basis. In this movie domain, we can have constraints that say - \textit{anyone who is a child should not be involved with discussions about R-rated movies} and this can be written as follows:
\begin{lstlisting}[language=prolog,basicstyle=\footnotesize]
    :- like_movie(P, M), is_adult_movie(M), age_category(P, children).
\end{lstlisting}

\noindent Similar to \textit{constraints},\textit{ preferences over defaults} are also an integral part of human reasoning process. An example rule will be: \textit{normally, people are more interested in talking about the actor than the director of a movie, however, people are more interested in talking about the director if he/she won an Oscar.}
\begin{lstlisting}[language=prolog,basicstyle=\footnotesize]
    movie_cast_crew_type(actor).     movie_cast_crew_type(director).
    like_director(P, D) :- movie(M), director(D), like_movie(P, M), movie_director
            (M, D), not ab_like_director(P, A).
    talk_preference(P, M, actor) :- actor(A), like_actor(P, A),
            not ab_talk_preference(P, M,actor), not neg_talk_preference(P,M,actor).
    talk_preference(P, M, director) :-  director(D), like_director(P, D), 
            not ab_talk_preference(P, M, director), 
            not neg_talk_preference(P, M, director).
    neg_talk_preference(P, M, X1) :- movie_cast_crew_type(X1),
            movie_cast_crew_type(X2), talk_preference(P, M, X2), X1 \= X2. 
    neg_talk_preference(P, M, director) :- actor(A), like_actor(P, A), 
            not ab_neg_talk_preference(P, M, director).
    ab_neg_talk_preference(P, M, director) :- director(D), like_director(P, D),
            director_award(D, oscar).
    ab_talk_preference(P, M, actor) :- director(D), like_director(P, D),
            director_award(D, oscar).
\end{lstlisting}

\noindent 
The above rules represent part of the knowledge that a socialbot must have to hold a conversation with a person (in addition to knowledge about movies from a movie-database such as iMDB). The socialbot will then compute the {\tt RCC($\tau$,$r$)} set given a topic {$\tau$} that interests the user ({\small {\tt like\_movie(john,titanic)}}) for example,  then select one of the {\small {\tt talk\_preference/3}} atoms to figure out what to say next. 
Given the topic {\small {\tt like\_movie(john, titanic)}}, the above rules with the iMDB database (restricted to data for 55 movies) computed the following RCC with a distance of 3 (we only show the {\small {\tt talk\_preference/3 predicates}} as those are most relevant and used for advancing the conversation: 

{\footnotesize 
{\tt $\{$talk\_preference(john,titanic,trivia), talk\_preference(john,titanic,awards),

~~talk\_preference(john,titanic,leonardo\_dicaprio)$\}$}
}

\noindent From this RCC set, we may pick {\small {\tt talk\_preference(john, leonardo\_di\_caprio)}}. Our algorithm will give us the ``path" in the dependency graph from {\tt like\_movie(john,titanic)} that leads to this inference. We will use the path to craft the response: ``Leonardo Di Caprio was the actor in the movie and did a great job." We may have also picked {\small {\tt talk\_preference(john, wolf\_of\_wall\_street, leonardo\_di\_caprio)}} and the path computed will tell us that the connection is through the lead actor De Caprio. So the socialbot will craft the response: ``Leonardo Di Caprio did a great job in Titanic. He also acted in the movie Wolf of Wall Street. Did you see that movie?". Phrases such a ``did a great job" and ``Did you see that movie?" are based on having further commonsense knowledge that associates such phrases with talking points ({\small {\tt talk\_preference/3}}) about an actor or a new movie mentioned.

Note that if we pick, for example, {\small {\tt talk\_preference(john,leonardo\_di\_caprio)}} then the fact {\small {\tt already\_discussed(john, titanic,leonardo\_di\_caprio)}} will be added to our database. When the RCC is computed for the next utterance,  we will not pick the topic {\small {\tt talk\_preference(john, titanic, leonardo\_di\_caprio)}} for advancing the conversation. 

{\footnotesize 
\begin{center}
\begin{table}[H]
\caption{Performance Table for DiscASP}
 \begin{minipage}{\textwidth}
 \begin{center}
    \begin{tabular}{lrrr}
\hline\hline
Query & Distance & \#Result & Execution Time (ms) \\
\midrule
like\_movie(john,titanic) & 3 & 3 & 9.90 \\
like\_movie(john,titanic) & 5 & 8 & 11.01\\
like\_movie(john,forrest\_gump) & 3 & 2  & 4.02 \\
like\_movie(john,forrest\_gump) & 5 & 2  & 7.61 \\
like\_actor(john,tom\_hanks) & 3 & 2  & 5.84 \\
like\_actor(john,tom\_hanks) & 5 & 6  & 5.04 \\
      \hline\hline
    \end{tabular}
    \end{center}
  \end{minipage}
\label{tb1}
\end{table}
\end{center}
}

\vspace{-0.25in} 

\noindent\textbf{Performance:} 
We use the rules from the movie example in Section \ref{sec:example} with a data set of 55 movies from iMDB movie database for performance testing of DiscASP system. As shown in Table \ref{tb1}, the queries are passed into the system, and each query has two different RCC distances (3 and 5), the outputs of the system are shown as the number of {\tt talk\_preference/2} atoms computed. The knowledge base has 740 grounded rules per user. As can be observed, query execution is quite fast. This is important in a conversational socialbot as the latency between the user utterance and the bot response cannot be too much (at most a few hundred milliseconds).

\section{Related Work} \label{sec:related work}

The CNR Dependency Graph and our DiscASP algorithm have a number of advantages: (i) partial answer set can be incrementally computed to find the RCC set; the chain of reasoning that connects one concept atom in the answer set to another can be found with no additional costs; this is in contrast to other methods such as CLINGO, where the \textit{entire} answer set has to be computed, and Galliwasp, where atoms related to proving a query only can be found. (ii) Declarative Knowledge: The DiscASP algorithm allows knowledge in a conversational socialbot to be represented declaratively as an answer set program, considerably reducing the complexity of the project. (iii) A better representation for Knowledge Graph: The CNR dependency graph provides a better representation for a knowledge graph, as it can account for exceptions and preferences as well. Traditional knowledge graphs such as ConceptNet \cite{conceptnet} and Microsoft Knowledge Graph \cite{mcg} represent knowledge by connecting words through relations and providing an ontology; they cannot represent exceptions and preferences. Thus, knowledge is inconsistent and wrong conclusions (such as ``penguins can fly") can be drawn. 

Logic programming has been used in the past for developing chatbots. Tarau and Figa  \cite{figa2004knowledge} represent  domain knowledge with propositional clauses in Prolog and find answers via using goal driven queries. 
Also, Tarau et al. \cite{tarau_blanco_2021} explored the use of dependency graphs with lexical knowledge and TextRank algorithm to give the most relevant answers given a query. 
In  recent times, most chatbots are developed using machine-learning technology \cite{chatbot-ref}. Researchers have developed many dialog datasets to train and test the language models, such as SwitchBoard corpus \cite{godfrey1992switchboard}, 
bAbI dialog dataset \cite{babi_dialog}, etc. 


These systems that are built with deep learning techniques learn the patterns of the training text remarkably well and show promising results on test data. With the recent advancements in the language model research, the pre-trained models such as BERT \cite{bert} and GPT-3 \cite{gpt3} show outstanding capability in generating natural language responses. Using these language models people have also developed techniques to build conversational chatbots, such as PD-NRG (policy driven neural response generator) \cite{PDNRG}. 
While these tools and techniques based on machine learning and deep learning are very impressive, they are based on syntactic and contextual pattern matching. They produce absolutely no understanding of the knowledge contained in the text. They utterly fail when any type of reasoning is required. Their black-box nature also precludes them from explaining their responses.   

There are also systems, such as StaCACK \cite{aaai21}, based on ASP that not only outperform the neural-models in accuracy on the bAbI-dialog dataset but also show natural language justification for each of its answers. Our conversational socialbot application using DiscASP is another step forward toward mimicking the human reasoning process in intelligent systems. We believe that, to obtain truly intelligent behavior, machine learning and commonsense reasoning should work in tandem.  

\section{Conclusion and Future Work} \label{sec:conclusion}

We proposed a dependency graph based algorithm called DiscASP to compute the atoms of an answer set that are related to a query atom within a fixed ``causal distance". We used a novel transformation to ensure that each program has a unique dependency graph, as otherwise multiple programs can have the same dependency graph. A major advantage of our algorithm is that it can produce a justification for any proposition that is entailed by a program. Currently, DiscASP only works for propositional answer set programs. Our goal is to extend it so that answer sets of datalog programs can also be computed without having to ground them first. This will be achieved by dynamically propagating bindings along the edges connecting the nodes in our algorithm's propagation phase. Thus, DiscASP will help us pave the way for finding answer sets efficiently without grounding the program first. The DiscASP algorithm is quite efficient and has been applied to practical problems such as developing conversational socialbots.

We described a real-world application of DiscASP to developing conversational socialbots. We showed how commonsense knowledge about holding a conversation about movies can be coded as an answer set program. We showed how atoms that are relevant to advancing a conversation given a topic can be found using the DiscASP algorithm. Future work includes incorporating knowledge about other social topics such as books, movies, sports, family, pets, etc., to realize a full-fledged socialbot using ASP technology. Current technology for developing socialbots is primarily based on machine learning. An effective socialbot can only be built if the socialbot ``understands" the utterances of the human user and can reason about them. Thus, knowledge-based approaches such as those based on ASP are crucial for developing conversational socialbots.





\medskip\noindent\textbf{Acknowledgment:}
Authors are supported by grants from NSF (IIS 1718945, IIS 1910131, IIP 1916206), Amazon and DoD.

\vspace{-0.2in}

\bibliographystyle{eptcs}
\bibliography{generic}
\end{document}